\documentstyle[sprocl]{article}

\newcommand\vev[1]{\langle{#1}\rangle}

\def\fun#1#2{\lower3.6pt\vbox{\baselineskip0pt\lineskip.9pt
        \ialign{$\mathsurround=0pt#1\hfill##\hfil$\crcr#2\crcr\sim\crcr}}}

\renewcommand\({\left(}
\renewcommand\){\right)}

\newcommand\eq[1]{Eq.~(\ref{#1})}
\newcommand\eqs[2]{Eqs.~(\ref{#1}) and (\ref{#2})}
\newcommand\eqss[3]{Eqs.~(\ref{#1}), (\ref{#2}) and (\ref{#3})}

\newcommand\ee{\end{equation}}
\newcommand\be{\begin{equation}}
\newcommand\eea{\end{eqnarray}}
\newcommand\bea{\begin{eqnarray}}

\newcommand\TeV{\,\mbox{TeV}}
\newcommand\GeV{\,\mbox{GeV}}
\newcommand\MeV{\,\mbox{MeV}}
\newcommand\keV{\,\mbox{keV}}

\newcommand\mpl{M_{\rm P}}

\newcommand\lsim{\mathrel{\rlap{\lower4pt\hbox{\hskip1pt$\sim$}}
    \raise1pt\hbox{$<$}}}
\newcommand\gsim{\mathrel{\rlap{\lower4pt\hbox{\hskip1pt$\sim$}}
    \raise1pt\hbox{$>$}}}

\newcommand\sub[1]{_{\rm #1}}

\newcommand\mgravvac{m_{3/2}}

\begin{document}

\title{CURRENT ISSUES FOR INFLATION}

\author{D. H. LYTH}

\address{Physics Department, Lancaster University, Lancaster
LA1 4YB, U.K.}

\maketitle\abstracts{Brief review of some current topics,
including  gravitino creation
and large extra dimensions.}

\paragraph{Most inflation models create a  lot of gravitinos}

I will focus on papers that appeared in 1999, 
building on a fairly comprehensive review \cite{treview}
of earlier work, and starting with gravitino creation.\footnote
{Updated version of a talk
given at COSMO99 International Workshop on Particle Physics and the 
Early Universe, 27 September--2 October 1999,  Trieste, Italy.}
Gravitinos are created at reheating  by 
thermal collisions \cite{subir}. If the gravitino mass $m_{3/2}$
is of order
$100\GeV$, as in gravity-mediated models of SUSY breaking,
these gravitinos upset nucleosynthesis unless  $\gamma T\sub R
\lsim 10^9\GeV$, where 
$T\sub R$ is  the reheat temperature, and
$\gamma^{-1}$ is  the increase in entropy per comoving volume (if any)
between reheating and nucleosynthesis. If instead $m_{3/2}\sim 100\keV$,
an in typical gauge-mediated models of SUSY breaking,
the gravitino is stable and will overclose the Universe unless
$\gamma T\sub R\lsim 10^4\GeV$. Only if $m_{3/2}\gsim 60\TeV$, as might
be the case in anomaly-mediated models of SUSY breaking, are the gravitinos
from thermal collisions completely harmless.
 
Gravitinos will also be created
  after inflation
\cite{lrs},  by the amplification of the
 vacuum fluctuation.
The evolution equations for the helicity $1/2$ and $3/2$  mode functions,
 required to calculate
this  second  effect, have been presented only this
year. A suitably chosen  helicity $3/2$ mode function satisfies 
\cite{mm} the Dirac equation in curved spacetime, with mass $m_{3/2}(t)$
(the gravitino mass in the background of the time-dependent
scalar field(s) which  dominate the Universe after inflation).
This implies that helicity $3/2$ gravitinos created from the vacuum
are cosmologically insignificant, compared with those created from
particle collisions \cite{lrs}.

The situation for helicity $1/2$ is more complicated, because this
state mixes with the fermions involved in SUSY breaking
(the super-Higgs effect). So far, the evolution equation for the mode
function has been presented
  \cite{kklv,grt} only for the
 simplest possible case, that 
 the only 
 relevant superfield is  a single chiral superfield.
Using this  idealized 
 equation, its  authors estimated (see also \cite{p99nont,p99new})
that   gravitinos
created {\em just after} inflation have,   at nucleosynthesis, the abundance
\be
\frac ns 
\simeq 10^{-2}\frac {\gamma T\sub R M^3}{V}
\,.
\label{ninf}
\ee
The abundance is specified by the ratio
of 
  $n$,  the gravitino number density, and $s$, the entropy density.
It is determined  by
 $V$, 
 the potential at the end of inflation, and 
  $M$, the mass
of the  oscillating  field which is  responsible for the energy density
just after inflation.

Entropy  increase can come from a late-decaying particle, with or 
without thermal inflation \cite{bg,thermal1,thermal2,thermal3}.
If there is no thermal inflation,
  the requirement 
 that final reheating occurs before nucleosynthesis gives
\be
\gamma T\sub R\gsim10\MeV
\label{gamtbound}
\,.
\ee
 One bout of thermal inflation
typically multiplies $\gamma$ by a  factor of order
 $e^{-10}\sim 10^{-15}$.

\eq{ninf} is not the end of the story. Rather, 
close examination \cite{p99new2} of the idealized mode function equation
reveals  that  gravitino creation continues until
$H$  falls below the true gravitino mass $m_{3/2}$.
This increases the   abundance to\footnote
{This late-time creation occurs only when SUSY is broken in the vacuum,
leading to a nonzero value for $m_{3/2}$. It occurs because global
supersymmetry then ceases to be a  good approximation, every time 
the potential dips through zero. The   models considered
in \cite{kklv,grt,p99nont,p99new} have unbroken SUSY in the vacuum,
so that global SUSY is a good approximation at all times, and helicity $1/2$
 gravitino
production becomes the same as Goldstino production. As is the case for
any spin $1/2$ particle, the production of the Goldstino ceases
soon after inflation ends.}
\be
\frac ns 
\sim  10^{-2}\frac {\gamma T\sub R M^3}{M\sub S^4}
\,,
\label{nvac}
\ee
where  $M\sub S=\sqrt{\mpl m_{3/2}}$ is the intermediate scale.
(The energy density  is of order $M\sub S^4$ when $H\sim m_{3/2}$.)

 The idealized mode function equation,
 used to obtain the above
results, assumes that the  superfield responsible for SUSY breaking in 
the vacuum 
is the same as the superfield(s) describing inflation.
This  will presumably not
be the case in reality. On the other hand,
 the non-adiabaticity responsible
for gravitino creation, present in the idealized case that these   two
superfields are identical, is unlikely to 
disappear  just because they are different.
Therefore,  \eq{nvac} should provide a reasonable estimate of the gravitino
abundance if reheating takes place after the epoch $H\sim m_{3/2}$.\footnote
{Just after inflation ends, a significant fraction of the 
energy of the oscillating field may  be drained off by
 preheating, into marginally relativistic bosons and/or
fermions. If this occurs, the idealized model will certainly be
invalidated for a while, but because
the new energy redshifts, and is anyhow never completely
dominant,  the idealized model is likely to  
become reasonable again after a few Hubble times. If so, it will
survive until reheating, defined as the epoch when practically
all of the oscillating energy is converted into thermalized
radiation.}
If, in contrast,  reheating  occurs earlier,
gravitino creation will certainly stop then because there is no
coherently oscillating field, and the  abundance will be
\be
\frac ns \sim 10^{-2} \gamma \(\frac M{T\sub R} \)^3\hspace{4em}
(T\sub R> M\sub S)
\,.
\label{nvac2}
\ee
Combining \eqs{nvac}{nvac2}, we see that 
{\em the maximal abundance occurs if $T\sub R\sim M\sub S$, with smaller
abundance if we either decrease  {\em or increase} $T\sub R$}.

In typical models of inflation and reheating, these
 gravitino abundances are huge compared with the abundance
from thermal collisions, and lead to far stronger constraints
on the $T\sub R $ and $\gamma$. Consider first the case of gravity-mediated
supersymmetry breaking, corresponding to $m_{3/2}\sim 100\GeV$
and $M\sub S \sim 10^{10}\GeV$. Then, nucleosynthesis requires
$n/s\lsim 10^{-13}$, and
\bea
\gamma &\lsim & 10^{-11} \(\frac{10^{10}\GeV}{T\sub R}\) \(\frac
{10^{10}\GeV}{M} \)^3\hspace{3em}
(T\sub R\lsim 10^{10}\GeV)\label{5} \\
\gamma &\lsim & 10^{-11} \(\frac{T\sub R}M\)^3
\hspace{3em} (10^{10}\GeV\lsim T\sub R)\label{6}
\,.
\eea
Alternatively, consider 
 the case of gauge-mediated SUSY breaking, with the favoured
values $m_{3/2}\sim 100\keV$ and $M_S\sim 10^7\GeV$. Then
the gravitino is stable, and the requirement that it  should
not overclose the Universe gives $n/s\lsim 10^{-5}$, and
\bea
\gamma & \lsim &10^{-3}\(\frac{10^7\GeV}{T\sub R}\)\(\frac{10^7\GeV}{M}\)^3
\hspace{3em}(T\sub R\lsim 10^7\GeV)\label{7}\\
\gamma & \lsim &10^{-3}\(\frac {T\sub R }M\)^3
\hspace{3em}(10^7\GeV\lsim T\sub R)\label{8}
\,.
\eea

These constraints,  are very strong in most  models
\cite{treview} of inflation.
For instance, the popular 
 $D$-term inflation model (and other models) requires  $V^{1/4}\sim M\sim
10^{15}\GeV$. Then, \eqss{gamtbound}{5}{6} require
  at least one bout of thermal inflation if SUSY-breaking is gravity-mediated.
	If instead it is gauge-mediated, \eqss{gamtbound}{7}{8}
 require
$T\sub R>10^{11}\GeV$, and again entropy production (though not necessarily
 thermal inflation).
 The only popular models where
the constraints are completely ineffective are those
with soft supersymmetry breaking during inflation, 
 leading to $M$ perhaps of order $\mgravvac$.
 Such models include modular inflation \cite{bg,banks99,beatriz99},
 and  hybrid inflation with soft supersymmetry breaking
(using a tree-level
 \cite{rsg} or loop-corrected \cite{ewanloop,grs} potential).

\paragraph{What sort of field is the inflaton?}
The rest of this review deals with various issues in inflation
model-building.
At the most primitive level, a model of inflation is simply a specification
of the form of the potential, but  one normally 
requires also that the form of the potential
looks  reasonable in  the context of particle physics.
In particular, one might be
 concerned if the  field values
 are big compared with
the ultra-violet cutoff $\Lambda\sub{UV}<M\sub f<\mpl$.\footnote
{The fundamental quantum gravity scale $M\sub f$ is  less than
the 4-dimensional Planck scale $\mpl$ if there are large extra dimensions.
We are, of course, talking  about the values of the 
{\em canonically-normalized} fields, 
with  the origin at a  fixed point of the symmetries.}
However, string theory gives us different  kinds of  scalar field.
There are, indeed,  the ordinary fields (matter fields)
 whose values should be
 small
compared with $\Lambda\sub{UV}$,
 if the form of the potential is to 
be under control.
Most models of inflation have been built with such fields  in mind,
though all too often one notices at the end of the calculation that
the magnitude of the inflaton field is at the Planck scale or bigger.

On the other hand, there are also
 moduli, 
which determine things like the gauge couplings and the size of extra
dimensions. String 
theory  can  give guidance about the form of their potential
 at field values of order $\mpl$, even if $M\sub f$ is much less than
$\mpl$ owing to the presence of large
extra dimensions. It is marginally flat enough to support inflation
\cite{bg,banks99},
a detailed investigation being necessary to  see whether
viable inflation occurs in a given model \cite{beatriz99}.
Yet more exotic fields might be contemplated. For instance, it
 has been suggested \cite{dt,d} 
that the inflaton corresponds
to the distance between $D$-branes, which are coincident now but were
separated at early times.
 The canonically normalized 
inflaton field is $\phi\simeq M\sub f^2 r$, where $r$ is the distance between
the branes and $M\sub f$ is the fundamental quantum gravity scale.
The regime $r\gsim M\sub f^{-1}$ presumably required by quantum gravity
now corresponds to $\phi$ {\em bigger} than
 $\phi\gg M\sub f$.  (From this viewpoint it is not clear how to justify
also the regime $\phi\ll M\sub f$, invoked in \cite{d}.)
 At present, we
do not know  which type
of model Nature has chosen. On the other hand, 
future measurements of the spectral
index will confirm or rule out  most of the 
forms of the inflationary
potential, that are natural in the context of  matter fields
\cite{treview}.

\paragraph{Hybrid inflation needs fairly large field values}
During  hybrid inflation, the slowly-rolling
 inflaton field $\phi$ couples to a second
field  $\chi$, holding the latter at the origin during inflation.
Ignoring loop corrections,
both $\phi$ during inflation, and the vev $\vev\chi$
 can be taken to be
very small on the Planck scale. 
Somewhat remarkably, it has been shown recently
  \cite{p99hyb} that this is no longer the case with the loop correction
included. For instance, it is found that in
the usual case that the hybrid inflation is supposed
to give the  primordial curvature  perturbation,
 $\phi$ during inflation
and/or $\vev{\chi}$ must be at least $10^9\GeV$.
 While far below the Planck scale, this
number is far {\em above} the electroweak scale. This means  that 
  hybrid inflation, with matter fields, cannot work
  in the context of TeV-scale quantum gravity.
Also, if  $\chi$ is  identified with
an electroweak  Higgs field, $\phi$ has to  be bigger
than $\mpl$, even if the curvature perturbation  comes from an
earlier era of inflation. This second result
 calls into question the viability
of  an otherwise attractive
model \cite{ewb}  of electroweak baryogenesis.

\paragraph{Extra dimensions}

 The growth industry this year has been 
the possibility that we live on a three-dimensional brane,
with $n\geq 1$ large   extra dimensions.
I will confine my remarks to  the case $n>1$ \cite{add},
because   the situation for the case $n=1$ \cite{hw,rs} is changing too 
rapidly so say anything useful.

It is assumed that Einstein gravity holds in the $4+n$ dimensions 
with some Planck scale $M\sub f$. To avoid obvious conflict with collider 
experiments one needs at least
$M\sub f\sim \TeV$, and this extreme case is the one that has received
the most attention.
With   $n>1$, and  the extra dimensions
stabilized, 
Einstein gravity  holds in our 4-dimensional spacetime
on scales bigger than the radius $R$ of the extra dimensions. The
4-dimensional  Planck scale  $\mpl$ is given by
$\mpl^2 \sim R^n M\sub f^{2-n}$. The thickness of our brane is presumably
 of 
order $M\sub f^{-1}$. Then, in  the regime where
 the $4+n$ dimensional  energy density 
is much less than $M\sub f^{4+n}$ (ie., well below the quantum gravity
scale) the energy density on our brane is  much
 less  than
$M\sub f^4$ \cite{kl}. Assuming that the extra dimensions are stabilized, the 
Hubble parameter in this  regime is   given by  $3H^2=\rho/\mpl^4\ll R^{-2}$.
We learn that, well below the quantum gravity regime, 
 Einstein gravity will  correctly describe the evolution of the
Robertson-Walker Universe, through the usual Friedmann equation
\cite{treview}.

While cosmological scales are leaving the horizon during inflation,
the extra dimensions must indeed be stabilized, since
significant variation would spoil the observed scale independence of the
spectrum of the primordial curvature perturbation.
The simplest hypothesis is that they remain stabilized 
thereafter, so that they have their present value while cosmological
scales leave the horizon. 
In that case, 
the mass of the inflaton {\em during inflation} (not necessarily in the
vacuum)  must be tiny, $m_\phi\lsim
M\sub f^2/\mpl$. This mass presumably requires protection from
supersymmetry \cite{p98tev}, but sufficient protection is problematic
because the inflaton has to communicate with the visible sector
so as to reheat, while in that sector the chiral supermultiplets have
 $\TeV$ mass splitting.
 Leaving aside that problem, new as opposed to hybrid inflation
may be quite viable \cite{bd,p99hyb}.  Another proposal \cite{dt}
is to use the field corresponding to the distance between D-branes,
though this does not seem to give a viable curvature perturbation.
 
An alternative  \cite{kl,adkm} is to  assume
  that while the curvature perturbation is generated,
 the extra dimensions
are stabilized, while
cosmological scales 
are leaving the horizon, 
with sizes much smaller than at present.
One  still needs  a second, short period of inflation to get
rid of the dangerous cosmological relics (moduli) associated with the
oscillation of the extra dimension about its present value.
(Indeed, it
 has been shown \cite{cgt} that when  entropy production finally ends,
the moduli must have their present
size, with an accuracy $10^{-14}(T\sub{R}/10\MeV)^{3/2}$.)
This  late inflation might be thermal \cite{d,p99hyb}
or  slow-roll \cite{halyo223}, thermal having the advantage that it allows
a bigger inflaton mass (though one that will  still
require protection from supersymmetry \cite{p99hyb}).

\paragraph{Some other recent work}

Many other papers on inflation have appeared in 1999.
Some of them address the problem of keeping the inflationary
potential flat, in the face of supergravity corrections.
For instance, \cite{cgr} presents a no-scale type model,
while several works \cite{assisted} pursue the paradigm
\cite{lms}  of assisted inflation.
There has been further consideration of  hybrid inflation
with a running mass \cite{grs}. 
Finally, a   completely new paradigm of inflation has been proposed
\cite{kinf}, in which the coefficient of the  kinetic term of the inflaton 
passes through zero.

\def\NCA{\em Nuovo Cimento}
\def\NIM{\em Nucl. Instrum. Methods}
\def\NIMA{{\em Nucl. Instrum. Methods} A}
\def\NPB{{\em Nucl. Phys.} B}
\def\PLB{{\em Phys. Lett.}  B}
\def\PRL{\em Phys. Rev. Lett.}
\def\PRD{{\em Phys. Rev.} D}
\def\ZPC{{\em Z. Phys.} C}
\def\PR{\em Phys. Rep.}
\def\RPP{\em Rep. Prog. Phys.}

\end{document}